\documentclass[reprint,amsmath,amssymb,aps,prb,superscriptaddress,showkeys]{revtex4-2}

\usepackage{silence}
\usepackage{hyperref}
\usepackage{lineno}
\usepackage{graphicx}
\usepackage{natbib}

\hypersetup{
	pdftitle={High harmonic generation along a structural phase transition in silicon},
	pdfauthor={D. N. Purschke et al.},
	pdfkeywords={high harmonic generation, semiconductor, nanotechnology, disorder, decoherence, attosecond, tunnel ionization, lightwave electronics, ion implantation, electron microscopy},
}

\newcommand{\um}{$\mathrm{\mu m}$}

\begin{document}

\title{High-harmonic generation along a structural phase transition in silicon}

\keywords{high-harmonic generation, semiconductor, nanotechnology, disordered solid, attosecond, tunnel ionization, lightwave electronics}

\author{D. N. Purschke}
\email{dpur@lle.rochester.edu}
\affiliation{Joint Attosecond Science Laboratory, National Research Council and University of Ottawa, Ottawa, Ontario K1N 5A2, Canada}
\affiliation{Laboratory for Laser Energetics, University of Rochester, Rochester, New York 14623, United States}

\author{D. Vick}
\affiliation{Quantum and Nanotechnologies Research Center, National Research Council, Edmonton, Alberta T6G 2M9, Canada}

\author{A. C\'ardenas}
\affiliation{Instituto de Ciencias de Materiales de Madrid, Consejo Superior de Investigaciones Cientificas, Madrid 28049, Spain}

\author{N. Haram}
\affiliation{Joint Attosecond Science Laboratory, National Research Council and University of Ottawa, Ottawa, Ontario K1N 5A2, Canada}

\author{P. Bastani}
\affiliation{Joint Attosecond Science Laboratory, National Research Council and University of Ottawa, Ottawa, Ontario K1N 5A2, Canada}
\affiliation{Advanced Laser Light Source, Institut National de la Recherche Scientifique, 1650 Boulevard Lionel-Boulet, Varennes, Qu\'ebec J3X 1P7, Canada}

\author{S. Gholam-Mirzaei}
\affiliation{Joint Attosecond Science Laboratory, National Research Council and University of Ottawa, Ottawa, Ontario K1N 5A2, Canada}

\author{S. Mokhtari}
\affiliation{Joint Attosecond Science Laboratory, National Research Council and University of Ottawa, Ottawa, Ontario K1N 5A2, Canada}
\affiliation{Advanced Laser Light Source, Institut National de la Recherche Scientifique, 1650 Boulevard Lionel-Boulet, Varennes, Qu\'ebec J3X 1P7, Canada}

\author{V. Jelic}
\affiliation{Joint Attosecond Science Laboratory, National Research Council and University of Ottawa, Ottawa, Ontario K1N 5A2, Canada}
\affiliation{Advanced Laser Light Source, Institut National de la Recherche Scientifique, 1650 Boulevard Lionel-Boulet, Varennes, Qu\'ebec J3X 1P7, Canada}

\author{J. Chen}
\author{J. Canlas}
\author{J. Tordiff}
\author{Md. W. Rahman}
\affiliation{Quantum and Nanotechnologies Research Center, National Research Council, Edmonton, Alberta T6G 2M9, Canada}

\author{A. Yu. Naumov}
\author{D. M. Villeneuve}
\author{A. Staudte}
\affiliation{Joint Attosecond Science Laboratory, National Research Council and University of Ottawa, Ottawa, Ontario K1N 5A2, Canada}

\author{M. Salomons}
\affiliation{Quantum and Nanotechnologies Research Center, National Research Council, Edmonton, Alberta T6G 2M9, Canada}

\author{R. E. F. Silva}
\affiliation{Instituto de Ciencias de Materiales de Madrid, Consejo Superior de Investigaciones Cientificas, Madrid 28049, Spain}
\affiliation{Max-Born-Institut for Nonlinear Optics and Short Pulse Spectroscopy, Berlin D-12489, Germany}
\author{\'A. Jim\'enez-Gal\'an}
\affiliation{Instituto de Ciencias de Materiales de Madrid, Consejo Superior de Investigaciones Cientificas, Madrid 28049, Spain}
\affiliation{Max-Born-Institut for Nonlinear Optics and Short Pulse Spectroscopy, Berlin D-12489, Germany}

\author{G. Vampa}
\affiliation{Joint Attosecond Science Laboratory, National Research Council and University of Ottawa, Ottawa, Ontario K1N 5A2, Canada}

% \linenumbers
\begin{abstract}
We explore high-harmonic generation (HHG) in silicon treated by a variable dose from a Ga focused ion beam (Ga$^+$ FIB). We identify four qualitatively different dose regimes as the sample transitions from a crystalline to amorphous phase. The phase transition manifests as a giant enhancement in yield of low-order harmonics, which is strongly correlated with surface swelling as measured by atomic force microscopy. Below the critical dose, we observe a mixed phase regime where the HHG spectra display elements of both crystalline and amorphous silicon. We show how the samples can be laser-annealed with intense non-resonant pulses, where energy is deposited preferentially into amorphous regions, allowing us to disentangle the relative importance of chemical and structural modifications to the strong-field dynamics. At very high dose, we observe an increased sensitivity to laser-induced damage, which is correlated with FIB-induced etching. Energy dispersive x-ray spectroscopy reveals that laser damage leads to the formation of Ga-rich nanoparticles and porosity at the surface. 
\end{abstract}

\maketitle

\section{Introduction}
\begin{figure*}
	\includegraphics{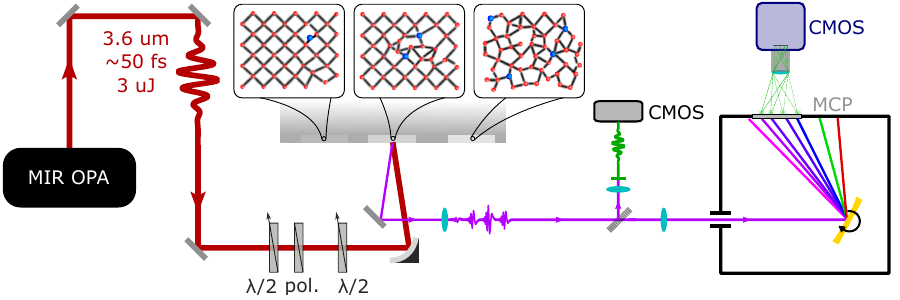}
	\caption{\label{fig:schematic} Schematic layout of the experiment. Pulses from a MIR OPA are focused onto a sample consisting of a FIB-treated silicon wafer, where they excite HHG that is routed to either a CMOS camera for sample alignment and real-time monitoring of damage/annealing or a UV spectrometer. The schematics above the sample illustrate the sample morphology in different regimes of Ga$^+$ dose, from left to right: defects and dopants, amorphous pockets in a c-Si matrix, and finally a continuous a-Si film. More experimental details are provided in the Methods section.}
\end{figure*}
In solid-state high-harmonic generation (HHG), an intense non-resonant laser field drives tunnel ionization of electron-hole pairs followed by free acceleration through the continuum. As in gases, the electron-hole pairs can recombine to emit high-energy photons at harmonics of the driving laser frequency \cite{vampa_linking_2015,vampa_theoretical_2014}, however, in contrast to atoms and molecules, the continuum is not the vacuum but the conduction and valence bands of the solids as dictated by the electronic structure \cite{yue_introduction_2022}. The electronic structure inherently modifies electron-hole trajectories, thus, it is imprinted into the spectrum of harmonic radiation. Today, HHG in crystals is a powerful spectroscopic tool sensitive to symmetry, topology, electronic structure, and crystal dynamics \cite{vampa_alloptical_2015,you_anisotropic_2017,langer_symmetry-controlled_2017,yoshikawa_high-harmonic_2017,jiang_crystal_2019,bionta_tracking_2021,nie_following_2023,takeda_ultrafast_2024,zhang_highharmonic_2024,heide_ultrafast_2024,zhan_rashba-induced_2026}. Furthermore, a variety of techniques from both attoscience and nanophotonics exist to control HHG in solids \cite{ciappina_attosecond_2017,korobenko_insitu_2022,purschke_microscopic_2023,roscam_abbing_enhancing_2024}.

However, while HHG has been observed in virtually all classes of materials \cite{liu_high-harmonic_2017,you_high-harmonic_2017,alcala_high-harmonic_2022,alexander_observation_2023,mondal_high-harmonic_2023,takeda_ultrafast_2024,shima_gholam-mirzaei_high_2025}, relatively little is known about the role of morphology and chemical composition in solid-state strong-field processes. Theoretical studies have suggested that HHG can be dramatically enhanced by doping \cite{huang_high-order-harmonic_2017,yu_enhanced_2019}, where mid-gap states are predicted to enhance the rate of tunnel ionization. However, experimental studies have revealed only minor enhancements in yield due to doping \cite{sivis_tailored_2017,nefedova_enhanced_2021}. Furthermore, HHG spectroscopy in solids has focused primarily on crystalline materials, with the exception of a select few studies. One prominent example is the pioneering demonstration of HHG in amorphous silica \cite{you_high-harmonic_2017}. More recent studies have begun to explore new directions bridging materials science and HHG, for example demonstrating switchable HHG in phase-change materials and liquid crystals \cite{korolev_tunable_2024,annunziata_high-order_2024}, or exploring how nanoscale morphology alters HHG spectra \cite{journigan_high_2024}. More generally, liquids, which bear close resemblence to amorphous solids, have recently been of interest in the attosecond science community \cite{alexander_observation_2023,mondal_high-harmonic_2023,mizuno_harmonic_2025,moore_solvation-induced_2025,mondal_multi-plateau_2026,tao_quantum_2026}.

Recently, we reported a detailed study of HHG in silicon samples amorphized by a Ga$^+$ focused ion beam (Ga$^+$ FIB) \cite{purschke_giant_2025}. FIB-driven amorphization of the Si surface leads to a  giant enhancement in tunnel ionization rate due to the reduction in band gap, enhancing low-order HHG yield by orders of magnitude. Simultaneously, disorder leads to a decoherence that quenches higher-order harmonics, lending new insight into the nature of dephasing in the solid state \cite{abadie_spatiotemporal_2018,floss_ab_2018,orlando_simple_2020,heide_probing_coherence_2022,brown_real-space_2024,cardenas_effects_2026,mokhtari_phonon-driven_2026}. Furthermore, the HHG spectra reveal hidden order inherited from the original crystalline phase and unseen by conventional spectroscopic probes. While ref. \citenum{purschke_giant_2025} focuses on the extremes (crystalline vs amorphous Si, a-Si vs c-Si, respectively), FIB enables continuous control over the Ga$^+$ dose. Measurements of Sivis et al. on low-order harmonics from FIB-patterned zone plates suggest a weak enhancement of h3 in the low-dose regime (Fig. S7 ref. \citenum{sivis_tailored_2017}). However, a broad range of conditions remain unexplored, leaving interesting questions, for example, into the nature of non-perturbative HHG and tunnel ionization in the mixed-phase regime or at very high dose.

The morphology of ion-implanted Si varies dramatically with dose and, due to its technological importance, has been the subject of extensive study for many decades \cite{howe_collision_1980,tamura_focused_1986,claverie_cross-sectional_1988,jones_systematic_1988,caturla_ion-beam_1996,benton_evolution_1997,basnar_focused_2003,pelaz_ion-beam-induced_2004,edmondson_amorphization_2009,drezner_high_2017,winkler_doping_2021,gibson_self-ion_2025}. For the purposes of the present work, we categorize the dose scale into 4 qualitatively different regimes. (1) Well below the critical dose, isolated collision cascades can create point defects and nm-scale amorphous pockets \cite{howe_collision_1980,jones_systematic_1988,caturla_ion-beam_1996}. With room temperature implantation, as in the present study, self-annealing during implantation tends to heal the small amorphous regions \cite{edmondson_amorphization_2009}, leaving vacancies and point and extended defects \cite{benton_evolution_1997}. (2) Increasing the dose, stable amorphous pockets begin to form slightly beneath the surface as collision cascades begin to overlap, leaving the sample in a mixed phase \cite{claverie_cross-sectional_1988,drezner_high_2017}. (3) Above the critical dose, the amorphous pockets percolate and form a continuous a-Si layer, such that the structure has completed the phase transition between crystalline and amorphous silicon (c-Si to a-Si) \cite{drezner_high_2017}. This regime is associated with swelling of the surface \cite{basnar_focused_2003,huey_low-dose_2003}. (4) Finally, at very high doses the Ga concentration becomes very large (percent scale) and the surface begins to etch away.

Here, we explore HHG through each of these regimes with an emphasis on the range over which the structural phase transition occurs. At the lowest dose, ion implantation weakly enhances perturbative nonlinearities of low-order harmonics. As the dose increases, the HHG spectra reveal a mixed-phase of a-Si and c-Si. In this regime, we demonstrate a type of femtosecond laser annealing with highly non-resonant excitation, where energy is deposited by tunnel ionization and thus preferentially targets a-Si regions. Crucially, annealing measurements allows us to isolate the relative importance of chemical vs structural modifications governing HHG in this system. Above the critical dose, where a continous amorphous layer is formed, the lower-order harmonic yield saturates. This saturation in HHG yield is strongly correlated with surface swelling as measured by atomic-force microscopy (AFM). At still higher dose, the FIB starts etching the surface, which is accompanied by a greatly reduced laser-induced damage threshold (LIDT). Finally, elemental maps of a lamella extracted from the laser-damaged surface in the high-dose regime hint at the formation of Ga-rich nanoparticles and enhanced surface porosity.

\section{Methods}
\subsection{HHG spectroscopy}
As shown in Fig. \ref{fig:schematic}, we use a mid-infrared (MIR) optical parametric amplifier (OPA, Light Conversion, Orpheus-MIR) outputting 3.6 \um{} pulses with approximately 50 fs duration and up to 3 $\mathrm{\mu J}$ pulse energy at 100 kHz. Further details of the system are provided in ref. \citenum{purschke_inlinedelay_2024}. The amplitude and polarization of the MIR pulses are controlled using a half waveplate/polarizer/half waveplate combination. Optical harmonics are generated from a silicon surface and collected in reflection mode. The harmonic radiation is focused into either a CMOS camera for sample alignment and real-time imaging of HHG yield or a spectrometer consisting of a UV grating (McPherson) and image intensifier with an S20UV photocathode (Photonis). The grating was rotated to capture spectra at different central wavelengths; three grating settings are required to capture the HHG spectrum from the 5$^{th}$ to 17$^{th}$ harmonics (h5-h17). The third harmonic was not measured as the intensifier is not sensitive to the 1200 nm wavelength. Additionally, the system is insensitive to h19 and higher ($\lambda \leq$190 nm) due to the $\sim 2$ m propagation length in air and UV-Al routing mirrors, however, the c-Si sample supports harmonic radiation at these higher orders \cite{suthar_role_2022}.

\subsection{Ion Implantation}
\begin{figure}
	\includegraphics{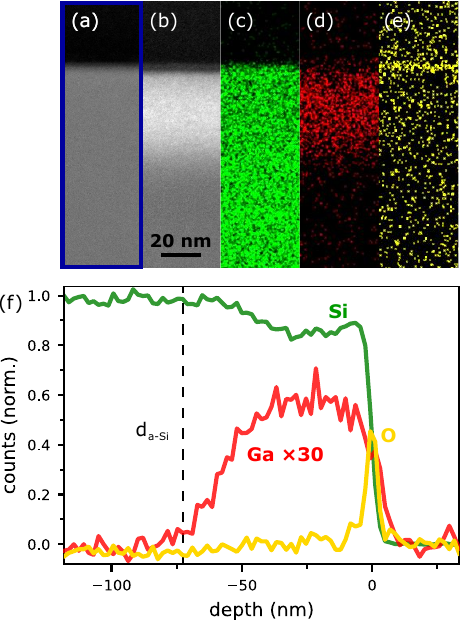}
	\caption{\label{fig:elemental_undamaged}Elemental depth profiling of a cross-section of the a-Si surface. (a-b) TEM images of lamellas in an unexposed region and exposed region, respectively, with Ga dose of $\mathrm{4\times 10^{16}\,cm^{-2}}$. The corresponding EDX elemental profiles measured at the (c) Si (green), (d) Ga (red), and (e) O (yellow) edges. (f) Depth profile taken from the images in (c-e).}
\end{figure}
Ion implantation was performed on a weakly n-type (carrier concentration $\mathrm{10^{16}\,cm^{-3}}$) silicon (100) substrate using a Zeiss NVision 40 dual beam microscope. 30 keV Ga$^+$ ions were focused onto a weakly n-doped silicon (100) substrate \cite{tamura_focused_1986,winkler_doping_2021}. An array of write fields of dimension 410 µm x 340 µm were exposed to create ion fluence conditions spanning 2$\times 10^{12}$ - $4\times 10^{16}$ $\mathrm{ions\cdot cm^{-2}}$ on a logarithmic scale, a sufficient range to explore each of regimes (1)-(4). The write fields were separated by 4 mm and marked with ion-milled fiducials at the write-field corners to facilitate ease of identification and alignment during the subsequent optical interrogation experiments. In order to produce ion fluence conditions spanning 4 orders of magnitude, two parameters were varied: the ion beam current (selected from 40 pA, 150 pA, 705 pA, and 6400 pA) and the total exposure time. One additional adjustment was done in order to ensure a laterally uniform exposure: the objective lens of ion column set to an underfocus condition to produce a beam spot diameter of 3 \um{}; this ensured that the ion beam overfilled the 186 nm pixel size of the raster pattern. All beam current settings were verified by means of a Faraday Cup prior to performing the exposures.

\subsection{Electron microscopy and Elemental mapping}
Site-selected lamella samples were prepared in a Hitachi NB5000 FIB. Prior to FIB milling of the lamella, the substrate was sputter coated with a carbon layer of thickness 40 nm. This step created a separation layer that allowed the Ga of the implantation layer to be readily distinguished from the additional Ga introduced during the preparation of the TEM sample in the FIB, which includes a gas-injection system (GIS) assisted 2 \um{} thick Ga-rich carbon coating. Complete amorphization of the sample at high dose, i.e., the destruction of long-range order and quenching of diffraction peaks, was verified using convergent beam electron diffraction (CBED). CBED was performed under scanning-TEM mode with a JEOL 2200FS TEM at an accelerating voltage of 200kV. The beam convergence angle for CBED pattern collection was 2.5 mrad and the electron beam waist was 1 nm at the focus.

Elemental maps were acquired from the lamellas on a JEOL JEM-ARM200CF transmission electron microscope equipped with a with large angle silicon drift detector energy dispersive x-ray (EDX) spectrometer. Fig. \ref{fig:elemental_undamaged}(a-b) shows transmission electron microscope (TEM) images of a lamella taken from untreated c-Si and a-Si. Fig. \ref{fig:elemental_undamaged}(c-e) show elemental maps of the a-Si sample measured at the Si, Ga, and O edges, respectively. Note that this sample had a Ga$^+$ dose of $\mathrm{4\times 10^{16}\,{cm^{-2}}}$, i.e., well above the critical dose to ensure a strong enough signal at the Ga edge in the EDX spectra. Fig. \ref{fig:elemental_undamaged}(f) shows the corresponding elemental profiles as a function of depth acquired by horizontally integrating the images in (c-e). The Ga signal, which was much smaller than both O and Si, was multiplied by 30x and is localized to the first 70 nm from the surface, which agrees with TEM measurements. The O signal indicates the oxidation layer at the surface. The reduced Si signal at the surface is likely explained by a combination of (1) increased electron backscatter and reabsorption of the Si-edge x-rays by the heavier Ga ions, (2) the reduction in density of a-Si due to amorphization, and (3) etching of the surface removing Si atoms. 

\begin{figure}[ht]
	\includegraphics{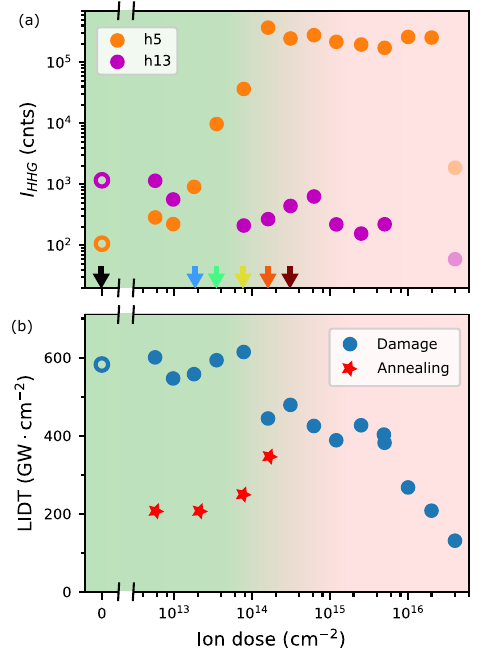}
	\caption{\label{fig:dose_scale} HHG from Ga$^+$ treated silicon. (a) Yields of h5 and h13 as a function of Ga$^+$ dose. These orders were chosen as representative of high- and low-order hamonics, respectively. The laser intensity was fixed to 210 $\mathrm{GW\cdot cm^{-2}}$. The colored arrows along the axis point to the doses at which the intensity scaling is plotted in Fig. \ref{fig:intensity_scaling}. The faded data points at $\mathrm{3\times 10^{16}\,cm^{-2}}$ indicate that the intensity was above the damage threshold for this sample.  (b) Laser-induced damage threshold (blue circles) as a function of ion dose and approximate intensity for the onset of laser annealing (red stars) for several samples. The colored backgrounds transitioning from green to red indicate the dose at which the sample transitions from crystalline to amorphous, as reported in the literature.}
\end{figure}

\section{Probing the structural phase transition with high-harmonic generation}
To begin, we study how the yield of the h5 and h13 changes as the structure is transformed with increasing Ga$^+$ dose, as reported in Fig. \ref{fig:dose_scale}(a). The peak intensity of the laser was fixed to $\mathrm{210\,GW\cdot cm^{-2}}$. At the lowest doses, up to $\mathrm{\sim 10^{13}\,cm^{-2}}$, we observe a small enhancement in HHG yield relative to untreated c-Si (consistent with Sivis et al., who reported a dose of $\mathrm{3\times 10^{12}\,cm^{-2}}$ \cite{sivis_tailored_2017}). In this regime, low-density lattice defects, dopants, and interstitials (regime 1) introduce mid-gap states that could enhance the perturbative nonlinearity of h5 or locally enhance tunnel ionization and non-perturbative HHG \cite{vampa_theoretical_2014}. Subsequently, the yield of h5 dramatically increases over the next decade of ion dose by 3 orders of magnitude due to the growth of low-bandgap a-Si pockets (regime 2). Note the c-Si direct band gap is 3.4 eV while in a-Si this is reduced to approximately 1.75 eV \cite{tauc_optical_1968,bose_electronic_1988}, which greatly enhances the tunnel ionization rate \cite{purschke_giant_2025}. In this regime, the measurement probes a macroscopic volume compared to the scale of a-Si pockets, thus, a mixed phase of a-Si and c-Si (this will be revisited in section \ref{sec:annealing}). 

At still higher doses, ranging from $2\times 10^{14}$ to $3\times 10^{16}\, \mathrm{cm^{-2}}$ the yield of h5 saturates. This coincides well with the critical dose near $\mathrm{10^{14}\,cm^{-2}}$ reported for complete amorphization using similar energy ions \cite{tamura_focused_1986} and implies that the saturation of yield is correlated with the formation of a continuous amorphous layer (regime 3). As shown in Fig. \ref{fig:dose_scale}(b), above the critical dose, an increasing Ga dose was also associated with a reduced laser-induced damage threshold. At very high dose (approaching $\mathrm{10^{16}\,cm^{-2}}$), the damage threshold begins to sharply drop off, marking the onset of regime 4, which will be revisited in section \ref{sec:damage}. Note that there was some variability in the LIDT for all treated samples.

\begin{figure*}
	\includegraphics{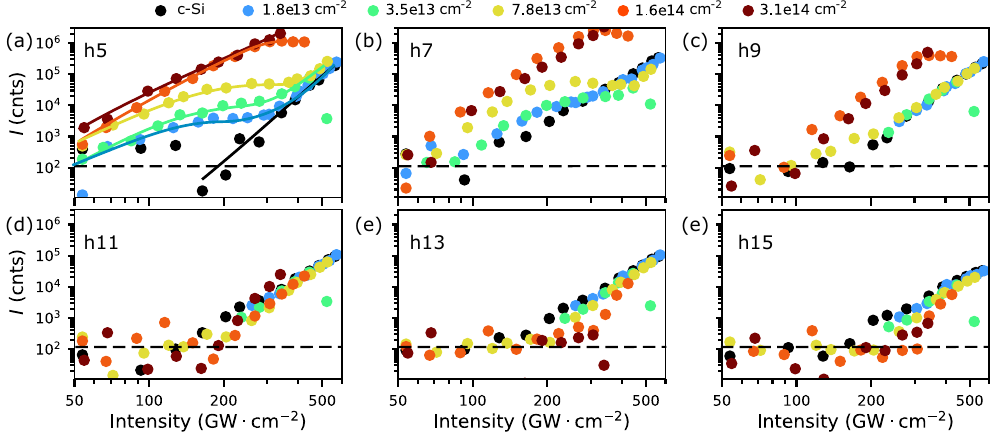}
	\caption{\label{fig:intensity_scaling} Intensity scaling along the structural phase transition. (a-e) Yields of odd-order harmonics from h5-h15, respectively, as a function of peak laser intensity for several ion doses. The solid lines in (a) are drawn to highlight the trend in yield with intensity. The laser intensity was increased from zero up to the visible damage threshold.}
\end{figure*}

Having identified regime (2), we can study how the nature of the strong-field response changes in the mixed-phase regime as amorphous pockets form and the structure transitions from crystalline to amorphous. Fig. \ref{fig:intensity_scaling}(a-e) reports the HHG yield for h5-h15, respectively, for several samples along the transition. All curves were measured starting at low intensity and increasing up to the visible damage threshold. We note that for the higher doses, there was a gradual increase in the scattered light background level in our spectrometer (not shown), making it more difficult to distinguish the weak high-order harmonics from the background.

Interestingly, the two samples with highest ion dose showed increased maximum yield of h5-h7 (Fig. \ref{fig:intensity_scaling}(a-b)) relative to c-Si at its own damage threshold. At higher dose, as shown in Fig. \ref{fig:dose_scale}(b) the LIDT is reduced, suggesting that there is an optimal dose for enhancing yield. The fifth harmonic in untreated c-Si (black circles, Fig. \ref{fig:intensity_scaling}(a)) demonstrates nearly perfect $I^5$ scaling expected for a perturbative nonlinearity, while higher order harmonics display clear non-perturbative scaling. Focusing on the highest-dose sample with $\mathrm{3.1\times 10^{14}\,cm^{-2}}$, which is above the critical dose, we see that the scaling law is altered by amorphization: h5/h11 display shallower/steeper power-law scaling than c-Si, respectively. The shallower scaling of h5 is associated with the reduced band gap, which decreases the Keldysh parameter so that the process becomes more non-perturbative. Conversely, the steeper scaling of high-order harmonics is associated with disorder-driven decoherence, which preferentially damps recollision-like processes for electron-hole pairs with longer trajectories through the continuum \cite{purschke_giant_2025}.

Apart from the highest two doses, all FIB treated samples display a shoulder in the yields of h5-h7 (Fig. \ref{fig:intensity_scaling}(a-b), respectively). While a shoulder in the intensity scaling of HHG from c-Si has previously been observed \cite{vampa_observation_2018}, the mechanism at play in FIB-treated Si is distinct. Below the shoulder, the yield of h5 for all treated samples is enhanced relative to c-Si, in agreement with the data in Fig. \ref{fig:dose_scale}(a). On the other hand, above the shoulder the curves of samples with doses up to $\mathrm{7.8\times 10^{13}\, cm^{-2}}$ collapse to nearly the same as c-Si. The next highest dose sample, with $\mathrm{1.6\times 10^{14}\, cm^{-2}}$, saturates in yield near the LIDT, however, due to the reduced damage threshold we were unable to reach the same peak intensity as c-Si. In h7, the shoulder was less apparent for the two lowest doses and in h9, only the sample with $\mathrm{7.8\times 10^{13}\, cm^{-2}}$ showed any hint of a shoulder. 

\begin{figure}
	\includegraphics{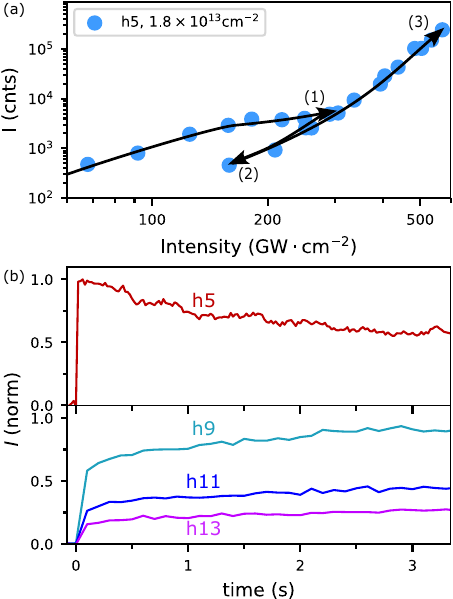}
	\caption{\label{fig:annealing_time}Irreversible modifications to HHG yield. (a) h5 measured as intensity (1) increases from 50-$\mathrm{300\,GW\cdot cm^{-2}}$, (2) is reduced from 300-$\mathrm{180\,GW\cdot cm^{-2}}$, and finally (3) increased up to the visible damage threshold. The arrows highlight the path taken. (b) Yield of several harmonic orders as a function of time after excitation in the sample with $\mathrm{3.5\times 10^{13}\, cm^{-2}}$. The laser repetition rate was 100 kHz and the acquisition time was 100 ms. The yield for h5 was measured on a CMOS sensor (20 ms acquisition time) with a bandpass filter while the higher-order harmonics were measured in the UV-vis spectrometer (100 ms acquisition time).}
\end{figure}

\section{Laser annealing in the tunnel ionization regime}\label{sec:annealing}
The collapse of intensity scaling curves from low-dose samples to that of c-Si suggests that the high-intensity laser anneals defects and amorphous pockets that exist below the critical dose. In Fig. \ref{fig:annealing_time}(a), we focus on the intensity scaling of h5 in the sample with an ion dose of $\mathrm{1.8\times 10^{13}\, cm^{-2}}$, i.e., in region (2), where the sample consists of a-Si pockets embedded in a c-Si matrix. The intensity was (1) increased from $\mathrm{50-300\,GW\cdot cm^{-2}}$, (2) reduced back down to $\mathrm{180\,GW\cdot cm^{-2}}$, finally (3) increased to the visible damage threshold. This path shows that the intense laser drives an irreversible change to the sample induced by exposure to the high-intensity laser without visible laser-induced damage. To explore the dynamics of this irreversible change, the sample was exposed to a laser with peak intensity of $\mathrm{410\, GW\cdot cm^{-2}}$, i.e., well above the shoulder in Fig. \ref{fig:annealing_time}(a), and the harmonic yield was measured in real time, as shown in Fig. \ref{fig:annealing_time}(b). Interestingly, the yield of h5 drops off gradually over a period of several seconds, while the yields of harmonics h9-h13 actually grow with time. 

\begin{figure}[htb]
	\includegraphics{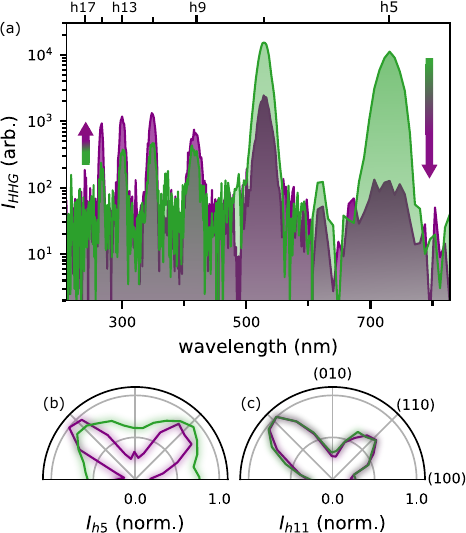}
	\caption{\label{fig:annealing}Laser annealing measured via spectra and polarization dependence. (a) Harmonic spectrum before (green) and after (purple) laser annealing. Both spectra were acquired with a peak laser intensity of $190\,\mathrm{GW\cdot cm^{-2}}$. In between acquisitions the sample was exposed to a higher intensity of $440\,\mathrm{GW\cdot cm^{-2}}$ for approximately 10 seconds. Yield of the (b) $5^{th}$ harmonic and (c) the $11^{th}$ harmonic as a function of laser polarization relative to the crystal axes of the bulk (unexposed) silicon crystal. The laser intensity was the same as for the spectra in (a).}
\end{figure}

HHG spectra acquired at a peak laser intensity of $\mathrm{190\,GW\cdot cm^{-2}}$ are shown in Fig. \ref{fig:annealing}(a) as measured before (green) and after (purple) exposure to the intense laser. This intensity was chosen to be below the shoulder in Fig. \ref{fig:annealing_time}(a). In further support of the laser-annealing hypothesis, we observe the opposite effect as that of amorphization \cite{purschke_giant_2025}, i.e., the lower-order harmonics are quenched and the higher-order harmonics are enhanced. Interestingly, focusing on the pre-annealed spectrum (green) we see characterstics of both a-Si and c-Si, namely, the enhanced low-order harmonic yield and the plateau in high-order harmonics, respectively. This is connected to the mixed-phase that exists as stable amorphous form, such that the plateau/lower-order harmonics are more sensitive to the c-Si/a-Si regions, respectively.

To emphasize that the intense laser is recrystallizing the structure, in Fig. \ref{fig:annealing}(b), we report the dependence of h5 on the pump beam polarization. In crystalline materials, pump-polarization dependence of HHG is sensitive to the crystal symmetry \cite{you_anisotropic_2017}, which, in (100) c-Si, manifests as sharply peaked harmonic yield when the laser is polarized along the (110) direction \cite{suthar_role_2022}. In FIB amorphized silicon, the lack of long-range order smears the sharp peaks out, however, the presence of residual medium-range preferentially aligned with the original crystal structure leads to slight polarization dependence in the harmonics \cite{purschke_giant_2025}. This structure is seen in Fig. \ref{fig:annealing}(b), which shows the polarization-dependence of h5 measured before and after laser annealing. Indeed, after laser annealing the polarization dependence develops the sharp peaks associated with c-Si, as shown by the purple curve in Fig. \ref{fig:annealing}(b). In contrast, the polarization dependence of h11 is essentially unaffected by annealing and, in fact, already displays the sharp peaks associated with c-Si before exposure to the annealing laser. Once again, this is connected to the mixed-phase that exists below the critical dose, where lower-order harmonics are sensitive to a-Si pockets while plateau harmonics are primarily sensitive to the c-Si regions.

These measurements allow us to disentangle the importance of chemical and structural modifications to the strong-field dynamics. Several recent theoretical and experimental studies have suggested that chemical modifications can be used to enhance HHG \cite{huang_high-order-harmonic_2017,yu_enhanced_2019,sivis_tailored_2017,nefedova_enhanced_2021}. In this work, upon annealing, which does not remove the implanted Ga, we observe a return to c-Si like HHG. This suggests that the structural modification is the dominant mechanism underlying the enhanced yield of low-order harmonics and quenching of high-order harmonics.

Although laser annealing in silicon has been studied extensively \cite{white_laserAnnealing_1979,sundaram_inducing_2002,mirza_non-thermal_2023}, along MIR driven amorphization for laser writing and surface modification \cite{werner_single-shot_2019,garcialechuga_amorphization_2024}, to our knowledge, this is the first demonstration of laser annealing in a-Si using a femtosecond laser in the tunnel-ionization regime. This regime could offer advantages over conventional techniques: the highly nonlinear tunnel-ionization process preferentially deposits energy into amorphous regions and defects. The nonresonant IR laser can transmit deep into a substrate before depositing energy into a target area, which could be useful, for example, as a tool for annealing buried structures \cite{nejadmalayeri_inscription_2005,chanal_crossing_2017,solana_silicon_2025} or  to directly write three-dimensional c-Si patterns in a-Si.

\section{Laser damage and etching}\label{sec:damage}
\begin{figure*}
	\includegraphics{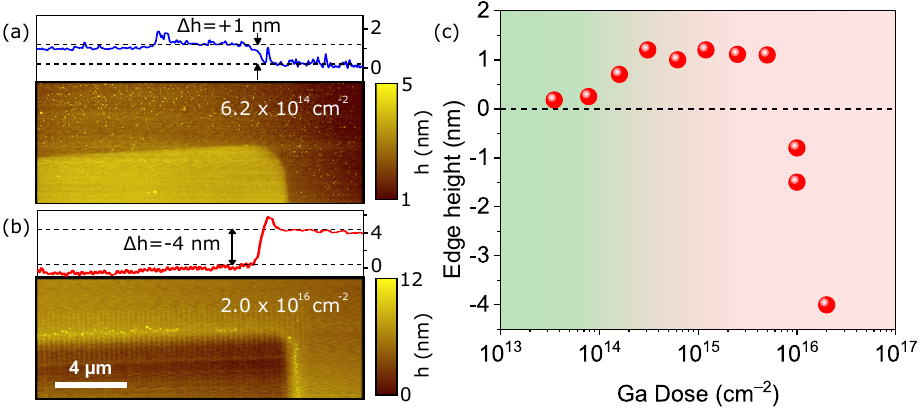}
	\caption{\label{fig:AFM}Atomic-force microscopy (AFM) images of FIB patterned a-Si regions with Ga doses of (a) $\mathrm{6.2\times 10^{14}\,cm^{-2}}$ and (b) $\mathrm{2\times 10^{16}\,cm^{-2}}$. The graphs above the images show lineouts along the step edge. (c) Step height as a function of Ga dose extracted from lineouts as in (a-b).}
\end{figure*}

Focusing now on the regimes above the critical dose, we find that the onsets of regimes (3) and (4), where we observed saturation of the h5 yield and, at higher dose, a rapidly dropping LIDT, respectively, are highly correlated with observations of swelling and etching via atomic force microscopy (AFM). As seen in Fig. \ref{fig:AFM}(a), at an ion dose of $\mathrm{6.2\times 10^{14}\,cm^{-2}}$, the surface shows a slight swelling of up to 1 nm due to the ion implantation. Further increasing the ion dose to $\mathrm{2\times 10^{16}\,cm^{-2}}$, the AFM image reveals FIB induced etching of the Si surface, as seen in Fig. \ref{fig:AFM}(b). 

We measured the step height using AFM for several samples above and below the cricial dose, as shown in Fig. \ref{fig:AFM}(c). The swelling gradually begins to grow at an ion dose around $\mathrm{10^{14}\,cm^{-2}}$ and peaks around 1 nm step height at a dose of $\mathrm{3\times 10^{14}\,cm^{-2}}$. Due to the low density of implanted ions relative to the Si, the swelling is associated with reduced density of a-Si relative to c-Si as opposed to the volume of implanted ions. From the 1 nm swelling, in combination with the 70 nm Ga implantation depth, we estimate a density reduction of 1-2 percent after amorphization. The saturation in swelling thus acts as an alternate measure of the critical dose, which coincides remarkably well with the plateau in the yield of h5 (Fig. \ref{fig:dose_scale}(a)).

Revisiting the LIDT (Fig. \ref{fig:dose_scale}(b)), above the critical dose and in the plateau region, there appears to be an abrupt drop in the LIDT, below which it is approximately equal to the damage threshold of c-Si. The abrupt drop occurs because at lower doses, laser annealing can heal the crystal structure and therefore mitigate the enhanced ionization resulting from amorphization. Near or above the critical dose, intensity required to begin annealing (red stars in Fig. \ref{fig:dose_scale}(b)) increased and, for samples above the critical dose, we were not able to rapidly anneal the samples. Though this does not preclude the possibility with different laser parameters, it does imply that a stable a-Si phase exists resilient to intense laser fields. The reduced damage threshold in a-Si seems to suggest that the a-Si is more susceptible to laser damage than c-Si, however, the more pertinent metric is LIDT relative to ionization fraction. Although ionization yield is difficult to measure experimentally, we note that other studies have linked it to HHG yield \cite{juergens_linking_2024}, which, as seen in Fig. \ref{fig:intensity_scaling}(a), is ultimately higher at LIDT for a-Si than c-Si. The degree of ionization at LIDT in a-Si is likely comparable or even greater than c-Si. The ultrathin nature of the film, where excitation is confined to the a-Si layer, limits the total deposited energy, which we speculate could be efficiently dissipated into the c-Si substrate to make the film more robust to laser damage.

At higher doses, above $\sim\mathrm{6\times 10^{15}\,cm^{-2}}$, the damage threshold begins to drop off rapidly. Here, the Ga concentration begins to approach the percent scale and the extremely high doping fraction could make the lattice less stable to non-thermal melting \cite{werner_single-shot_2019}. As shown in Fig. \ref{fig:AFM}(c), AFM reveals that this higher-dose regime is also correlated with the onset of etching at a Ga dose around $\mathrm{10^{16}\,cm^{-2}}$. We observed an approximately linear etch rate with dose. At the highest doses, the reduced damage threshold made it much harder to generate HHG. Clearly, an optimal Ga$^+$ dose exists for creating stable a-Si that can support the maximum possible laser intensity. Our observations suggest that this is near the critical dose, however, more experimentation is necessary to explore the trade offs between resistance to annealing and stablity to laser damage.

\section{Surface modification}\label{sec:elemental}
\begin{figure}
	\includegraphics{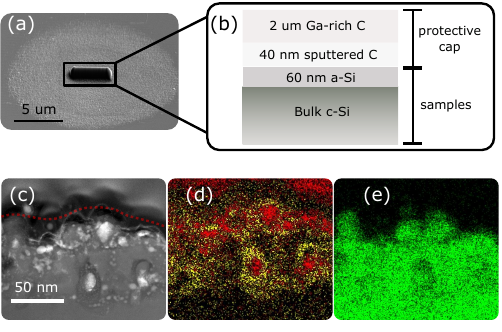}
	\caption{\label{fig:elemental_damaged}Ga-rich nanoparticle formation. (a) SEM image of the laser damaged spot. The black ridge in the middle is the capping layer deposited before lamella milling. (b) Layer structure of the lamella highlighting the protective cap and sample layers. (c) TEM cross section of the laser-damaged site. The red dashed line shows the approximately location of the carbon protective cap. EDX elemental profiles measured at (d) the Ga (red) and O (yellow) edges and (e) at the Si (green) edge.}
\end{figure}

Finally, in the high dose regime, we study the effect of laser damage on the morphology and chemical composition of the surface. A second lamella was prepared using the technique described in the methods section. An SEM image taken prior to milling is shown in Fig. \ref{fig:elemental_damaged}(a) for a laser-damaged region of the sample with an ion dose of $4\times 10^{16} \mathrm{\,cm^{-2}}$. To aid in interpretation, Fig. \ref{fig:elemental_damaged}(b) shows a schematic of the resulting lamella including the layers deposited for protection. 

Fig. \ref{fig:elemental_damaged}(c) shows a TEM cross-section of the laser-damaged sample lamella and, in contrast to the smooth interface of the undamaged samples, damage manifests as a highly irregular surface. Here, the red dashed line is drawn to approximately deliniate the sputtered carbon capping layer from the rough surface. Near the surface, there appears to be agglomeration into nanoparticles with varying brightness in the TEM image. Fig. \ref{fig:elemental_damaged}(d) shows an overlay of the EDX images measured at the O (yellow) and Ga (red) edges while Fig. \ref{fig:elemental_damaged}(e) shows the image at the Si edge (green). Interestingly, the bright regions in the TEM image correspond to Ga-rich nanoparticles. Furthermore, the oxygen signal so deep in the sample suggests that the laser damage induced porosity into the surface. As seen in Fig. \ref{fig:elemental_undamaged}(a), this buried O was not present in the undamaged sample.

\section{Conclusion}
We explored HHG along a structural transition between crystalline and amorphous phases in silicon driven by ion implantation with a Ga$^+$ FIB. We identified 4 qualitatively different regimes as a function of ion dose: (1) at low dose, a small enhancement in low-order harmonic yield due to mid-gap states from defects and dopants, (2) at higher dose, amorphous pockets form and dramatically enhance HHG as they grow, (3) a stable amorphous layer forms and the low-order harmonic yield saturates, and (4) at high implantation density, the structure becomes more susceptible to laser damage, which is correlated with the onset of FIB etching. In regime (2), we demonstrated a type of laser annealing where energy is deposited via tunnel ionization that preferentially targets a-Si regions and defects, which could have practical applications for annealing buried structures, while in regime (3) a stable a-Si layer with the current laser parameters was formed. Importantly, this also allows us to disentangle the role of structural and chemical changes in governing the HHG yield. Finally, we showed that in the high-dose regime, laser-induced damage leads to the formation of Ga-rich nanoparticles and porosity at the surface. Future work could explore more of the broad parameter space of ion-treated silicon, for example, using different ions and accelerating voltages or using thermal annealing to relax the structure. More generally, our work points towards new applications of HHG as spectroscopic tool in materials science.

\section{Acknowledgements}
DNP acknowledges funding from the Natural Sciences and Engineering Research Council of Canada (NSERC) Postdoctoral Fellowship Program and the Mitacs Globalink Research Award program. GV acknowledges funding from NSERC Discovery Grant program and the Joint Center For Extreme Photonics. AJG acknowledges support from the Comunidad de Madrid through TALENTO Grant 2022-T1/IND-24102 and the Spanish Ministry of Science, Innovation and Universities through grant reference PID2023-146676NA-100. The authors wish to acknowledge technical support from R. Kroeker and D. Crane.

\bibliographystyle{apsrev4-2}
\bibliography{HHG_structural_phase.bib}

\end{document}